\documentclass[10pt]{article}

\usepackage[OE]{express}
\usepackage{graphicx}

\begin{document}
	
	\title{Large-format platinum silicide microwave kinetic inductance detectors for optical to near-IR astronomy} 

	\author{P.~Szypryt,\authormark{1,*} S.~R.~Meeker,\authormark{1} G.~Coiffard,\authormark{1} N.~Fruitwala,\authormark{1} B.~Bumble,\authormark{2} G.~Ulbricht,\authormark{1,3} A.~B.~Walter,\authormark{1} M.~Daal,\authormark{1} C.~Bockstiegel,\authormark{1} G.~Collura,\authormark{1} N.~Zobrist,\authormark{1} I.~Lipartito,\authormark{1} and B.~A.~Mazin\authormark{1}}
	
	\address{\authormark{1}Department of Physics, University of California, Santa Barbara, California 93106, USA\\
		\authormark{2}NASA Jet Propulsion Laboratory, Pasadena, California 91109, USA\\
		\authormark{3}Dublin Institute for Advanced Studies, 31 Fitzwilliam Place, Dublin 2, Ireland}
	
	\email{\authormark{*}szypryt.paul@gmail.com}
	
	\date{\today}
	
	\begin{abstract}
		We have fabricated and characterized 10,000 and 20,440 pixel Microwave Kinetic Inductance Detector (MKID) arrays for the Dark-speckle Near-IR Energy-resolved Superconducting Spectrophotometer (DARKNESS) and the MKID Exoplanet Camera (MEC). These instruments are designed to sit behind adaptive optics systems with the goal of directly imaging exoplanets in a 800--1400~nm band. Previous large optical and near-IR MKID arrays were fabricated using substoichiometric titanium nitride (TiN) on a silicon substrate. These arrays, however, suffered from severe non-uniformities in the TiN critical temperature, causing resonances to shift away from their designed values and lowering usable detector yield. We have begun fabricating DARKNESS and MEC arrays using platinum silicide (PtSi) on sapphire instead of TiN. Not only do these arrays have much higher uniformity than the TiN arrays, resulting in higher pixel yields, they have demonstrated better spectral resolution than TiN MKIDs of similar design. PtSi MKIDs also do not display the hot pixel effects seen when illuminating TiN on silicon MKIDs with photons with wavelengths shorter than 1~$\mu$m.
	\end{abstract}

	\noindent
	Users may use, reuse, and build upon the article, or use the article for text or data mining, so long as such uses are for non-commercial purposes and appropriate attribution is maintained. All other rights are reserved.
	
	
	\providecommand{\noopsort}[1]{}\providecommand{\singleletter}[1]{#1}%

	\section{Introduction}
	
	Over the past three decades, Low Temperature Detectors (LTDs) have found a breadth of new applications in a number of fields, such as astronomy and beamline science. These detectors operate by exploiting various superconducting phenomena, and they can be separated roughly into two groups. The first group consists of thermal detectors, which operate by measuring temperature changes in an absorbing material due to incident photon power. Transition-Edge Sensors (TESs~\cite{Irwin1996, Irwin2005}) are the most common detectors in this group.  Their principle of operation involves biasing the detectors at the superconducting transition temperature, an area with a high temperature coefficient of resistance, resulting in high sensitivity to photon energy and therefore resolving power. Alternatively, there are more exotic thermal detectors such as metallic magnetic calorimeters (MMCs~\cite{Fleischmann2005, Fleischmann2009}), which use the dependence of magnetization on temperature to do this measurement.
	
	Athermal, or non-equilibrium, detectors make up the second group of LTDs. These detectors work by measuring the number of quasiparticles that are generated when a photon strikes a superconductor and breaks Cooper Pairs. The energy of the incident photon is proportional to the number of quasiparticles generated, giving these detectors spectral resolution. Examples of athermal LTDs include Superconducting Tunnel Junctions (STJs~\cite{Peacock1996}) and Microwave Kinetic Inductance Detectors (MKIDs~\cite{Day2003, Zmuidzinas2012}). MKIDs are described in considerable detail below.
	
	LTDs have a few chief advantages over more conventional charge-coupled devices (CCDs). The superconducting bandgap is roughly $10^4$ times smaller than that of silicon, the material typically used in CCDs. This lowers the radiation energy detection threshold as compared with CCDs, making these detectors extremely useful for submillimeter astronomy and the cosmic microwave background (CMB). For photon energies well above the bandgap, a large number of quasiparticles are created for single photon hit events, allowing for high spectral resolving power. In addition, LTD readout schemes can often be designed to bypass read noise, making these detectors ideal for observing faint sources.
	
	One of the LTD technologies listed above that has been proven to be particularly useful for ultraviolet, optical, and near-IR (UVOIR) astronomy is the MKID. MKIDs exploit a superconducting thin film's large kinetic inductance~\cite{Mattis1958} to enable highly sensitive photon detection. When a photon hits a superconductor it breaks up a number of Cooper pairs proportional to the energy of the incident photon. The broken Cooper pairs temporarily generate a number of quasiparticles, and because there are now less charge carriers in the superconductor, the inductance is proportionally increased. By devising a scheme to quickly measure this inductance shift, one can extract the energy of the photon that caused the shift and the time of arrival of the photon.
	
	MKIDs utilize superconducting LC resonators to do this sort of measurement. When a photon strikes the inductor portion of a resonator it raises its inductance.  The resonant frequency of the resonator is analogously decreased, as the LC resonator frequency goes as $1/\sqrt{LC}$. In order to read out the resonator it is coupled capacitively to a microwave transmission line and a probe tone is sent through the line to continuously drive the resonator and monitor any frequency shifts due to photon absorption events. 
	
	MKIDs are naturally multiplexed by coupling many resonators to the same microwave transmission line and using a digital readout system~\cite{McHugh2012}.  This readout sends a comb of probe tones through the transmission line at the resonant frequencies of all the resonators on that line.  The same digital readout system can be used to simultaneously monitor any photon events on these resonators. We currently read out two thousand resonators with a single microwave transmission line~\cite{Strader2016b}. The architecture of the readout electronics make it easier to monitor the phase shift response at the probe tone frequencies rather than to track the shifting resonator frequencies.  The energy of the photon is linearly proportional to the change in phase, but begins to become nonlinear when the phase shift signal approaches half a resonator line width, causing a degradation of spectral resolution for these high photon energies.  The sensitivity of the detectors, defined as the phase shift for a given photon energy, needs to be carefully designed so that photon energies detected by the instrument stay mainly within the linear regime.
	
	
	The first astronomical UVOIR MKID instrument, initially fielded at the 200" Hale Telescope at Palomar Observatory in 2011, was the Array Camera for Optical to Near-IR Spectrophotometry (ARCONS~\cite{Mazin2013}). ARCONS contained a 2024 resonator MKID array with substoichiometric titanium nitride (TiN) as the resonator material~\cite{Leduc2010}. The resonators were split between two coplanar waveguide (CPW) transmission lines, and the arrays were read out using eight Reconfigurable Open Architecture Computing Hardware (ROACH~\cite{Parsons2006}) digital electronic boards. Resonators were spaced roughly 2~MHz with gaps every 253 resonators to distinguish between resonator groups read out by different ROACH boards. The resonators were designed to fit in a microwave readout range of 3--6~GHz. The array was optimized for detection of photons between 380--1150~nm, with a maximum quantum efficiency of $\sim$17\%. The spectral resolution, $\lambda$/$\Delta\lambda$, was 8 at 400~nm.
	
	Although ARCONS was a technological breakthrough for UVOIR MKIDs, there were a number of detector issues that needed to be addressed before the technology could be used for the next generation of astronomical instrumentation. The main issue was the non-uniformity in composition in the substoichiometric TiN, which led to resonances appearing away from their designed values. This caused resonator collisions in frequency space, rendering many resonators unusable by the readout and lowering the overall usable detector yield to $\sim$70\%. 
	
	Another significant issue was related to the stability of pixels in the array.  When ARCONS was illuminated with photons above the bandgap energy of silicon (1.1~eV), pixels were occasionally observed to exhibit a temporary (timescale of seconds to minutes) telegraph noise in the phase direction that mimicked incident photons at a very high count rate~\cite{vanEyken2015}.  We believe that this was due to electrons that have been excited to the silicon conduction band interacting with the electromagnetic field of the resonator.  Using a higher bandgap substrate such as sapphire has been shown to eliminate this effect.
	
\begin{figure}[ht]
		\centering
		\includegraphics[width=0.7\linewidth]{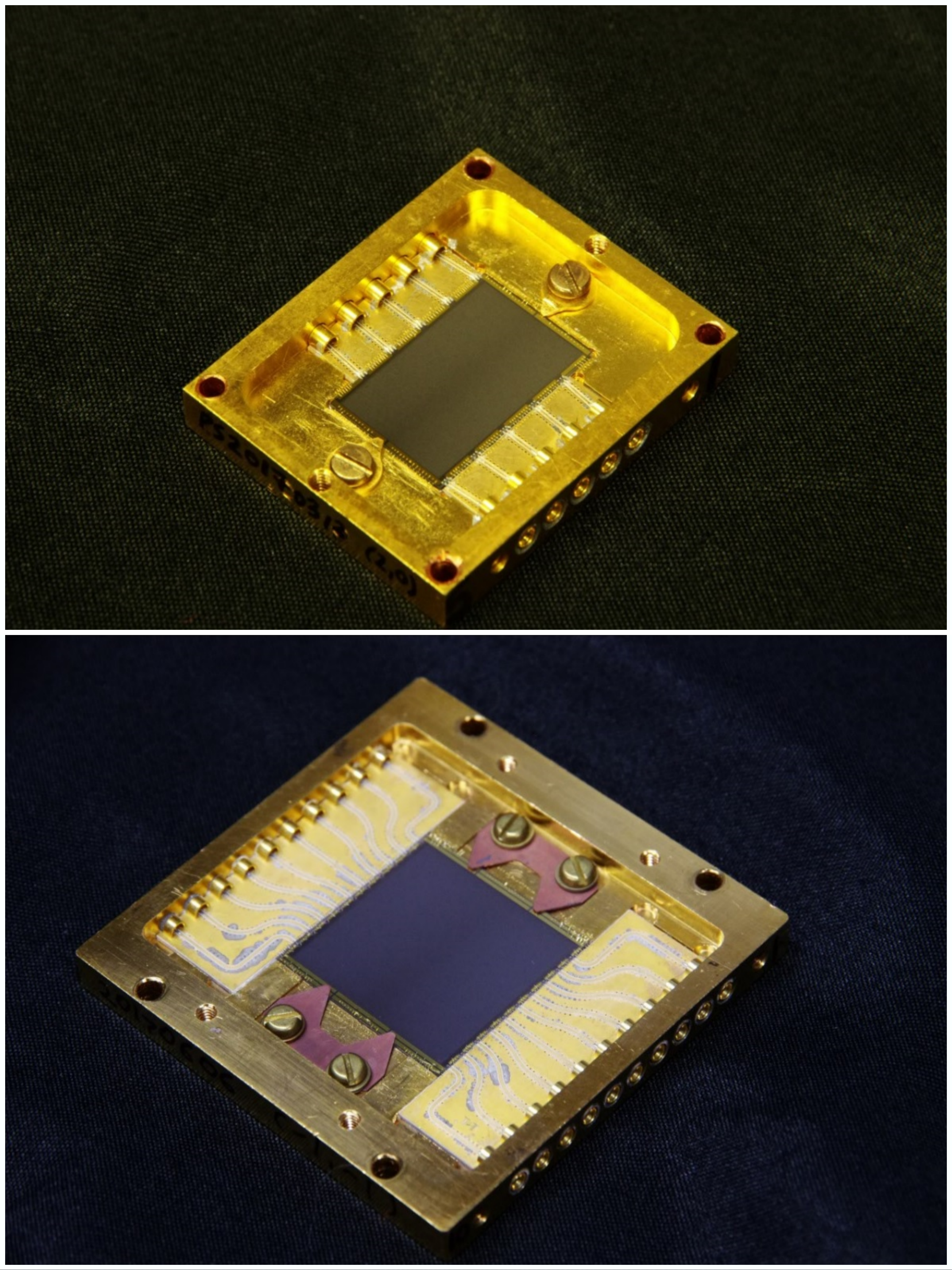}
		\caption{DARKNESS (top) and MEC (bottom) arrays after they have been mounted in their sample boxes and wire bond connections have been made. The size of the DARKNESS array is 13.5~cm~$\times$~20.3~cm whereas the MEC array is 24.6~cm~$\times$~22.5~cm.}
		\label{fig:DeviceBoxes}
	\end{figure}

	Here we present our work in addressing these issues and developing large-format arrays for two upcoming MKID instruments designed for the direct imaging of exoplanets~\cite{Meeker2015}. The first is the Dark-speckle Near-IR Energy-resolved Superconducting Spectrophotometer (DARKNESS). This instrument has already been commissioned at Palomar Observatory, but detector improvements continue to be made and arrays can easily be interchanged between observing runs. The second of these instruments is the MKID Exoplanet Camera (MEC), which is planned to be commissioned by the end of 2017 on the Subaru Telescope. Images of MKID arrays used in the DARKNESS and MEC instruments are shown in Fig.~\ref{fig:DeviceBoxes}.
	
	\section{Design}
	
	Large-format MKID arrays designed for various instruments share many common structures. These structures include superconducting LC resonators, transmission lines, a ground plane, and bond pads.  The superconducting resonators act as the photon detectors.  The transmission lines, which are usually made with coplanar waveguides (CPW) or microstrips, are used to drive and read out resonators and a single ground plane provides a common ground potential for the array.  Superconducting bond pads are used to connect the transmission lines to the device, and gold (Au) pads are used to remove excess heat from the chip to the device box with Au bond wires.  We typically make our device boxes out of gold-plated oxygen-free copper.
	
	\subsection{Superconducting Resonators}
	
	The most critical structures in the MKID array design are the superconducting resonators. A superconducting resonator consists of a meandered inductor that acts as the photosensor and an interdigitated capacitor (IDC) with legs of variable lengths to adjust the frequency. There are a number of design constraints that limit the choice of material and geometry of the resonators, and a summary of important resonator parameters is shown in Table~\ref{tab:design}. The critical temperature (T$_C$) should be kept as low as possible in order to maximize the sensitivity of the detectors and reduce background noise level\cite{Gao2008b}. The operating temperature of the refrigerator, however, also provides a lower limit on the T$_C$. Our current adiabatic demagnetization refrigerators (ADRs) are required to hold their temperature for an entire night of astronomical observations, limiting the operating temperature to $\sim$100~mK. More specifically, we achieve hold times of up to 14 hours when operating at 100~mK. We have found that the resonator performance saturates at operating temperatures of around T$_C$/8, requiring us to choose materials with a T$_C$ at or above 800~mK. Superconducting films with a T$_C$ between 0.8--1~K are ideal for astronomical UVOIR MKID instruments based on the current generation of ADRs.
	
	\begin{table*}[ht]
		\caption{DARKNESS and MEC Resonator Design Parameters}
		
		\begin{center}
			\vspace{0.5cm}
			\begin{tabular}{ l l }
				\multicolumn{1}{p{6.5cm}}{\textbf{Common Material Properties}} &
				\multicolumn{1}{p{5.75cm}}{ } \\
				
				\hline
				
				Resonator Material & PtSi \\
				Pt : Si Atom Number Ratio & 1.0 : 1.2 \\
				Superconducting Critical Temperature & 900~mK \\
				Film Thickness & 50~nm \\
				Surface Inductance & 10.5~pH/$\square$ \\
				
			\end{tabular}
			
			\vspace{0.25cm}
			
			\begin{tabular}{ l l l }
				\multicolumn{1}{p{6.5cm}}{\textbf{Other Design Parameters}} &
				\multicolumn{1}{p{2.75cm}}{\textbf{DARKNESS}} & 
				\multicolumn{1}{p{2.5cm}}{\textbf{MEC}}\\
				\hline
				Inductor Width $\times$ Height & 47~$\mu$m~$\times$~32~$\mu$m & 47~$\mu$m~$\times$~35~$\mu$m \\
				Average Inductor Leg Width & 1.7~$\mu$m & 1.7~$\mu$m \\
				Inductor Slot Width & 0.3~$\mu$m & 0.5~$\mu$m \\
				Interdigitated Capacitor Width $\times$ Height & 121~$\mu$m~$\times$~92~$\mu$m & 125~$\mu$m~$\times$~92~$\mu$m \\
				Interdigitated Capacitor Leg Width & 1~$\mu$m & 1~$\mu$m \\
				Interdigitated Capacitor Slot Width & 1~$\mu$m & 1~$\mu$m \\
				Frequency Band & 4.0~--~8.2~GHz & 3.8~--~8.0~GHz \\
				Average Coupling Quality Factor & 30,000 & 30,000 \\
				
			\end{tabular}
		\end{center}
		\label{tab:design}
		
	\end{table*}
	
	Another important resonator parameter is the internal quality factor (Q$_i$).  This is essentially a measure of the any losses in the superconducting resonator. Q$_i$ values of good resonators are typically >$10^5$, but values of over $10^6$ have been produced in TiN~\cite{Leduc2010}. High values of Q$_i$ appear to be correlated with better resonator performance due to higher maximum readout power and lower noise levels, which results in improved spectral resolution. Another source of power loss in the resonator is due to the coupling of the resonator to the CPW transmission.  This is measured through the coupling quality factor (Q$_c$), which is typically engineered to have a value of $\sim$30,000. The total resonator quality factor (Q$_r$) goes as (1/Q$_i$ + 1/Q$_c$)$^{-1}$, so when Q$_i$ is very high, Q$_r$$\approx$Q$_c$. The detector sensitivity is tuned assuming this condition, so having high Q$_i$ on the majority of resonators is also important for sensitivity uniformity across the array.
	
	The surface inductance of the superconducting film and inductor geometry also play an important role in designing the resonators. The fractional change in broken Cooper Pairs for a given photon energy scales with the total inductor volume, so this quantity needs to be carefully engineered to provide the optimal sensitivity for detecting photons within the energy band of interest. For a given resonator design this quantity can only be changed in fabrication through the thickness of the superconducting film, which is roughly inversely proportional to the surface inductance for thin films. We generally design the inductor volume and sheet impedance first, which sets the total amount of inductance within the resonator.  The capacitance can then be designed to place the resonator at a desired frequency. The DARKNESS and MEC arrays are optimized to detect photons within the 800--1400~nm wavelength band. The inductor volume is tuned so that the most energetic photons in this wavelength span produce roughly 120$^\circ$ phase pulses, which is near the onset of nonlinearity.
	
	For the DARKNESS and MEC arrays in this work we have decided to replace the TiN films that were used in the ARCONS arrays with platinum silicide (PtSi).  This material has many of the desired quantities listed above, such as a T$_C$ of 900~mK and quality factors of up to $10^6$, as was seen with single layer test devices~\cite{Szypryt2016}. This material also shows roughly an order of magnitude better uniformity than TiN ($<3\%$ deviation in sheet resistance across a 4" wafer), allowing for much more precise resonator frequency placement (designed for 2~MHz spacing) and therefore a higher usable detector yield. PtSi also has roughly a factor of two higher quantum efficiency than TiN at the wavelengths of observation of DARKNESS and MEC. The superconducting films were thick enough so that the optical absorption was limited by reflections off of the film surface rather than transmission through the film. In making the switch from TiN to PtSi, we also switched from using a silicon substrate to sapphire.  Although this was mainly done to get around a film growth issue that was lowering the Q$_i$ of the PtSi films, the change was also expected to reduce hot pixel behavior due to the decreased number of long-lived excited electrons in the semiconducting silicon substrate. It should also be noted that submillimeter MKID arrays could get around the uniformity issue by using aluminum (Al)~\cite{Goupy2016}, but this material is too reflective at UVOIR wavelengths and would severely lower quantum efficiency. Al also has a relatively low kinetic inductance fraction, which would reduce detector sensitivity.
	
	\subsection{Transmission Lines, Ground Plane, and Coupling Bars}
	The next set of structures that are important for driving the resonators are the transmission lines, the ground plane, and the coupling bars. In our current large-format array design, these structures are all made out of superconducting niobium (Nb) with a T$_C$ of $\sim$9~K and surface inductance of $\sim$0.1~pH/$\square$. The transmission lines take a fairly standard CPW design, with a width of 7.5~$\mu$m and slot size of 1.5~$\mu$m. This is done to ensure that the CPW lines are closely impedance matched to 50~$\Omega$ across the microwave bandwidth used by the readout. These transmission lines wind across the array multiple times in order to couple power to each individual resonator. This causes the neighboring ground planes to become electrically disconnected. In order to combine these various sections and form one cohesive ground plane, crossovers need to be incorporated into the fabrication process, as is discussed in Sections~\ref{sec:fabrication}.B--D.
	
	The coupling bars (capacitors) are used to couple power between the transmission lines and the resonators. These appear beneath the IDC portion of each resonator and also require crossovers to go over the ground plane and reach the resonator from the transmission line. The lengths of these coupling bars are adjusted to the frequencies of the resonators to which they are coupling power.  This is done to adjust the coupling strength to the individual resonators and set the Q$_C$ at 30,000, making the sensitivity and power handling more constant across the array.
	
	\subsection{Bond Pads}
	The final relevant structures in the large-format MKID array design are the bond pads. First, there are Nb bond pads at the ends of the Nb CPWs. These are a few hundred microns in length and width, and the slot size between the bond pads and the ground plane is increased to match the width to slot size ratio elsewhere in the transmission line.  This allows for straightforward Al wire bonding to these bond pads while keeping the transmission line matched as closely to 50~$\Omega$ as possible.  Wire bonds lengths are kept to a minimum to reduce reflections. There is also a set of Au bond pads covering the perimeter of the array. These are used to make Au bond wire connections from the box to the chip to remove heat from the array (superconducting Al wires have poor heat conductivity). Au wire bonds only adhere to other Au surfaces, which is why Au is necessary for these particular bond pads.
	
	\subsection{DARKNESS and MEC Design Variations}
	
	The DARKNESS and MEC arrays have very similar designs with the major difference being that the MEC array contain roughly twice the number of pixels as the DARKNESS array. The DARKNESS array has a total of 10,000 (125$\times$80) pixels across 5 CPW transmission lines. Each transmission line drives a subarray of 2,000 (25$\times$80) pixels that is read out using 2 ROACH2 digital electronics boards. These boards are the second version of the boards used for the ARCONS array and have improved bandwidth and resources~\cite{Strader2016b}. A total of 10 ROACH2 boards are required to read out the entire DARKNESS array. The MEC array is designed for a total of 20,440 (140$\times$146) pixels. These pixels are separated into 10 transmission lines and a total of 20 ROACH2 boards are required to read out the entire array. Each transmission line drives 2,044 (14$\times$146) resonators in this case. The array size is largely limited by the cost of the readout electronics and spacing between resonators. Although decreasing this spacing would theoretically increase the total number of resonators that could fit within a readout board's bandwidth, it would also lead to a higher number of frequency collisions that lower percent pixel yield. With these trade-offs in mind, a resonator spacing of 2~MHz was chosen.
	
	The MEC array also has some additional design improvements over the DARKNESS array.  One particular improvement is increasing the spacing and varying the distance of the crossovers connecting various segments of the ground plane by arbitrary amounts. This was done to reduce the impact of standing waves in the CPW line and improve the transmission through the line. The inductors of the MEC array also have gaps of 500~nm rather than the 300~nm that were used in the DARKNESS design. The larger gaps ensured that the inductors would be etched completely during the final fabrication step without over-etching into other areas of the resonator.  This was expected to reduce variations in frequency in the resonators and potentially raise Q$_i$. Additionally, the MEC fabrication mask included gaps around the coupling bars. This was done to re-etch the coupling bars in one of the final fabrication steps and ensure exact spacing between the coupling bar and the resonator and ground plane, giving better control of the Q$_C$. It also was done to eliminate any possible shorts between the coupling bars and the ground plane to increasing feedline yield.
	
\begin{figure*}[ht]
		\centering
		\includegraphics[width=1.0\linewidth]{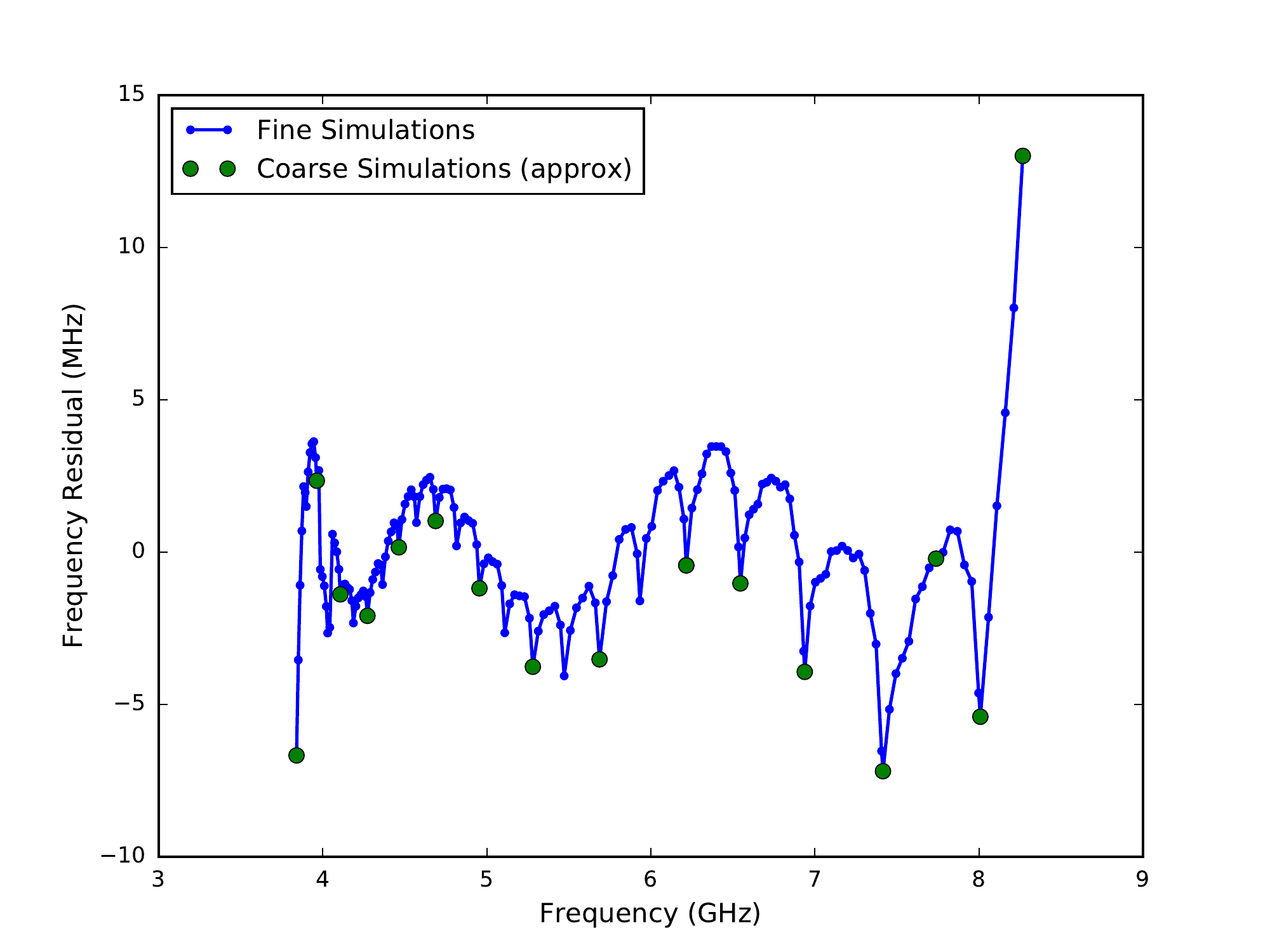}
		\caption{Resonators were simulated at 185 different capacitor leg lengths. The resulting frequency versus leg length curve was fit to a sixth order polynomial and the residuals of this fit as a function of frequency are shown in the plot above. A similar fit (with 15 simulations, shown approximately by the green points) was used to place frequencies on the DARKNESS array, resulting in errors of up to 15 MHz. In addition to the large scale sinusoidal feature not captured by the fit, most of the old simulations were done for total capacitor leg lengths where each leg was either fully extended or fully contracted (legs are varied one at a time to achieve a desired total capacitance). The variation in the frequency versus leg length curve as an individual leg is grown/shrunk was not accounted for (these are the humps between the green points). For the MEC array design, capacitor leg length was mapped to frequency by interpolating between fine simulations (blue points).}
		\label{fig:FreqSimPlot}
	\end{figure*}

	Finally, the MEC array resonator frequency simulations were redone to improve the precision of the 2~MHz resonator spacing across the 4~GHz bandwidth of the array. For the DARKNESS array, 15 resonators with different capacitor leg lengths were simulated, and the resulting resonant frequency versus leg length curve was fit to a 6th order polynomial. This fit was used to map resonant frequency to capacitor leg length for the final array design. Prior to the MEC array design process, it became clear that 15 simulations did not adequately sample the leg shrink versus frequency curve. This method was particularly unable to capture jumps in frequency that occurred when adjusting the lengths of two successive IDC leg pairs, resulting in errors of up to 15 MHz. To remedy this, a new set of simulations with 185 different capacitor leg lengths was used. A finer meshing setting was also used to improve simulation accuracy. Mapping leg length to resonant frequency was done by linearly interpolating between these simulations rather than using a higher-order polynomial fit. All simulations were done using Sonnet. A comparison of these two methods is shown in Fig.~\ref{fig:FreqSimPlot}.
	
	\section{Fabrication}
	
	\label{sec:fabrication}
	
\begin{figure*}[ht]
		\centering
		\includegraphics[width=\linewidth]{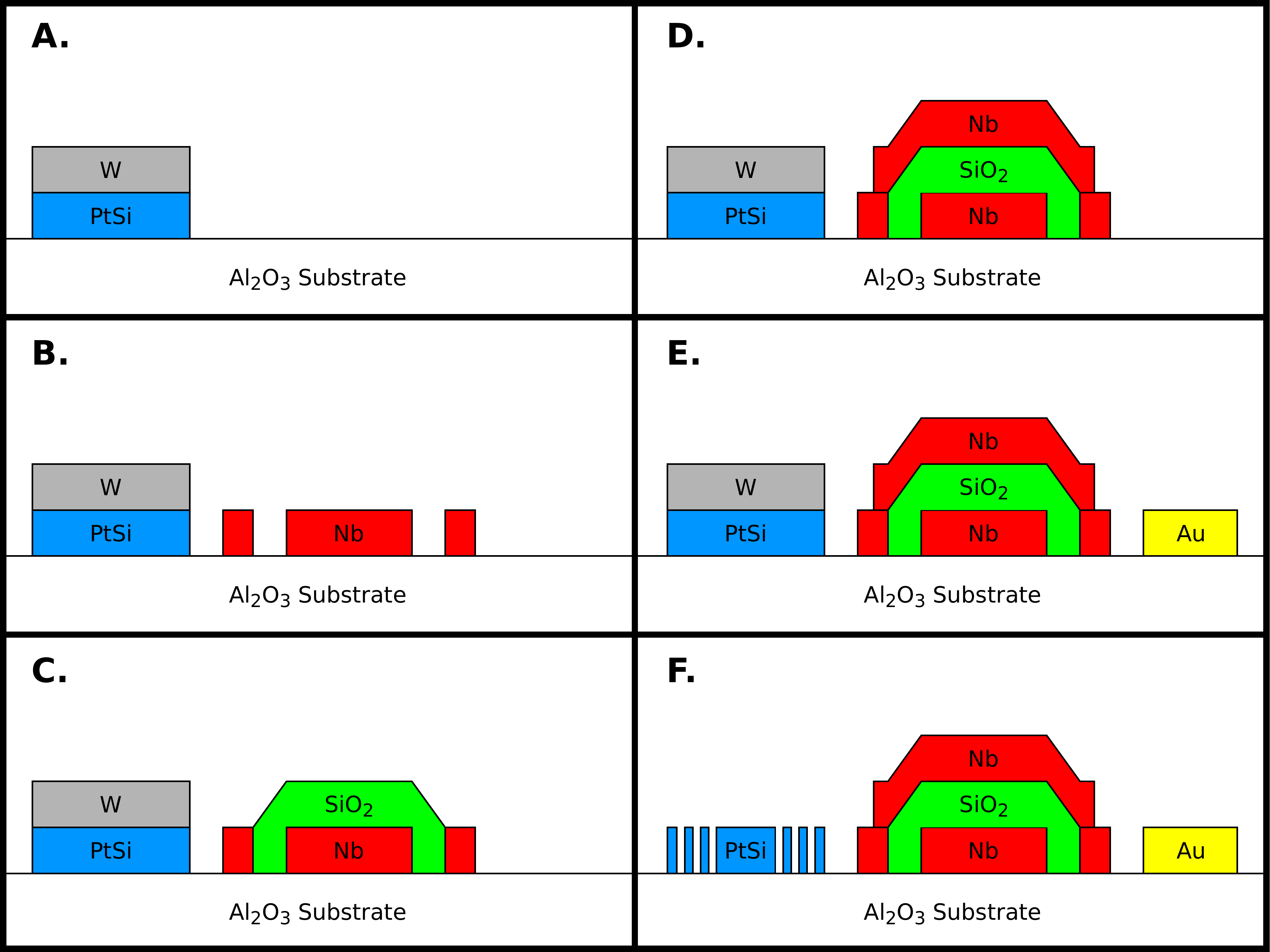}
		\caption{Fabrication cross-sections. A. Box outlines etch. B. Transmission lines and ground plane lift-off. C. Dielectric pads lift-off. D. Crossovers and coupling bars lift-off. E. Gold bond pads lift-off. F. Resonators etch.}
		\label{fig:cross-sections}
	\end{figure*}

\begin{figure*}[ht]
		\centering
		\includegraphics[width=\linewidth]{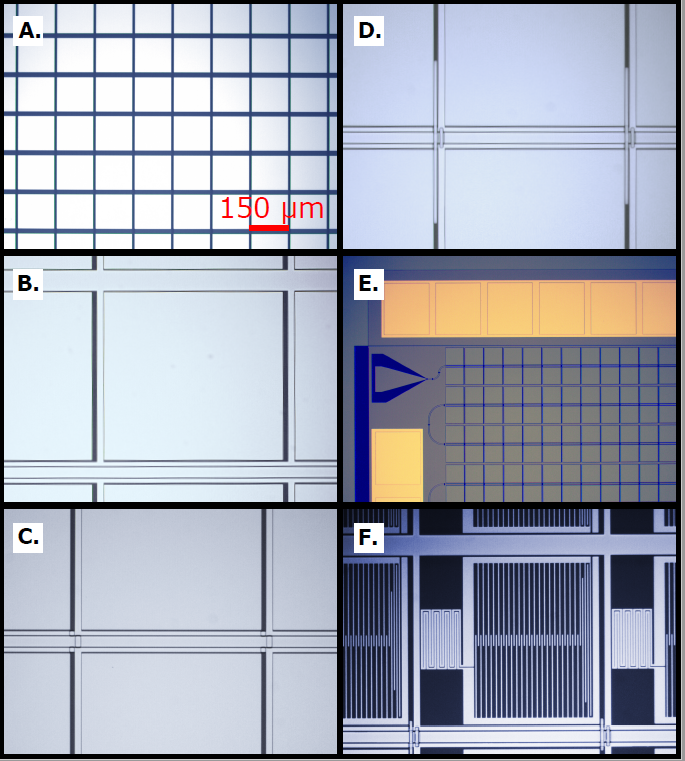}
		\caption{D3 microscope images at the end of each step. Images in the panels have been taken at different magnifications. For a sense of scale, the pixel pitch (distance between boxes or resonators) of the array is 150~$\mu$m. A. Box outlines etch. B. Transmission lines and ground plane lift-off. C. Dielectric pads lift-off. D. Crossovers and coupling bars lift-off. E. Gold bond pads lift-off. F. Resonators etch.}
		\label{fig:D3Images}
	\end{figure*}

	The fabrication of the large-format PtSi MKID arrays in this work can be divided into six steps corresponding to the six lithographic layers in the design.  MKID arrays of various sizes can be fabricated using identical processing steps, with the main difference being the photolithographic mask used to define the structures on the array. The fabrication of individual layers is described in detail below.  Fabrication cross-sections are shown in Fig.~\ref{fig:cross-sections}. Microscope images taken at each layer of the fabrication of a 10,000 pixel DARKNESS array are shown in Fig.~\ref{fig:D3Images}.
	
	\subsection{Box Outlines Etch}
	
	The first step in the large-format MKID array fabrication process is the deposition of the superconducting material that will be used for the sensor layer. To begin, a c-plane single crystal sapphire wafer is solvent cleaned and aggressively oxygen plasma cleaned using a Gasonics Aura 2000 plasma asher. In the current design, a $\sim$50~nm PtSi film is prepared on the sapphire substrate. The deposition and annealing are done in an AJA sputter system with a cryogenic pump, enabling the main chamber to reach a base pressures in the upper $10^{-10}$~Torr. A detailed description of the sputtered PtSi films used to make high internal quality factor MKIDs is given in Ref.~\cite{Szypryt2016}. In the case of large-format arrays a 100~nm tungsten (W) film is deposited directly on top of the PtSi film in the same sputter system without breaking vacuum between the two depositions.  This W film is used to protect the sensitive PtSi film from contamination and other fabrication issues until the final step.
	
	After deposition, a dual layer adhesion promoter and positive deep UV photoresist is spun onto the PtSi film. A box outline layer is defined in the photoresist using an ASML 5500 deep UV stepper capable of performing 150~nm lithography. This deep UV stepper is used in all subsequent photolithography steps. AZ 300 MIF is used for development. After lithography and development, the PtSi+W film is etched using a Panasonic E640 inductively coupled plasma (ICP) etcher. First, the W is etched using SF6. Next, the PtSi is etched using a chemistry of 60\%~Ar, 20\%~CF$_4$, and 20\%~Cl$_2$. The remaining photoresist is removed with solvents and a moderate oxygen plasma.
	
	The end result of this layer is shown in the microscope image in Fig.~\ref{fig:D3Images}(A). This step leaves boxes of PtSi in the areas that will ultimately contain the resonators. The areas that gets etched away will be used for the transmission line, ground plane, coupling bars, and bond pads.
	
	\subsection{Transmission Lines and Ground Plane Lift-off}
	
	The second layer in the fabrication process involves defining the coplanar waveguide transmission lines and ground plane segments in the areas where the PtSi got etched away in the previous step. To begin, a dual liftoff layer and positive photoresist are spun onto the wafer. The resist is exposed with the stepper and developed. To prevent liftoff artifacts, the wafer is run through a gentle oxygen plasma cleaning directly before the deposition.
	
	Next, a 90~nm Nb film is deposited using the same sputter system that was used for the PtSi deposition. The Nb is lifted off using n-methylpyrrolidone (NMP) and the wafer is cleaned for the next layer using solvents and a moderate oxygen plasma. At the completion of this layer, the Nb transmission lines and segments of ground plane will be defined on the wafer between boxes of PtSi, as shown in Fig.~\ref{fig:D3Images}(B) The next two layers are done to connect the various segments of these ground planes without shorting the transmission lines and to define capacitor structures that couple power from the transmission line to the detectors.
	
	\subsection{Dielectric Pads Lift-off}
	
	In the third fabrication layer, dielectric material is deposited and patterned to form insulating pads to facilitate superconducting crossovers. Fig.~\ref{fig:D3Images}(C) shows how these dielectric pads look on a DARKNESS device before the deposition of the crossovers. In this step, photoresist is spun onto the wafer and once again patterned using the stepper.  A 180~nm silicon dioxide (SiO$_2$) film is deposited onto the wafer by reactively sputtering from a Si target in an O$_2$ atmosphere. To do this, a different AJA sputter system with oxygen access, but higher base pressure, is used. The SiO$_2$ is lifted off using NMP and the wafer is once again cleaned to prepare it for the crossover layer.
	
	More recently, amorphous silicon (a-Si) has been used in place of SiO$_2$ for the dielectric layer due to its low expected loss tangent~\cite{Oconnell2008, Coiffard2017}. Here, a 180~nm thick a-Si film is deposited onto the wafer with plasma-enhanced chemical vapor deposition (PECVD) in a SiH$_4$ and Ar atmosphere. This is done with a UNAXIS VLR system, which performs the PECVD at higher gas densities than traditional PECVD systems. This results in a dense, high quality film with especially low dielectric loss tangent\cite{Oconnell2008}.
	
	\subsection{Crossovers and Coupling Bars Lift-off}
	
	The next layer in the fabrication process is the deposition and patterning of Nb strips over the dielectric pads deposited in the previous layer. This is done to connect various segments of the ground plane over the transmission lines in order to create a single, contiguous ground plane.  In addition, Nb bars are patterned below the capacitors of each resonator and are connected to the transmission line with a different set of crossovers. These bars are used to couple power from the transmission line to the resonators.
	
	This layer is once again started with spinning and patterning of photoresist using the DUV stepper. Nb is sputtered onto the wafer with the same conditions as were used for the ground plane deposition in the second layer, except here a longer deposition is used to produce a thicker, 300~nm film. The increased thickness of this film is necessary to avoid step coverage issues at locations where the Nb slopes over the dielectric pads. After deposition, the Nb is lifted off in NMP and cleaned, defining the structures shown in Fig.~\ref{fig:D3Images}(D).
	
	\subsection{Gold Bond Pads Lift-off}
	
	In the final lift-off layer, Au is deposited and patterned into bond pads around the perimeter of the MKID array. Au bond pads are necessary for thermally connecting the array to the box via Au wire bonds, allowing heat to escape the chip. The Au bond pads from the corner of a DARKNESS array are shown in Fig.~\ref{fig:D3Images}(E).
	
	A Temescal VES-2550 e-beam evaporator is used for the deposition. First, a 5~nm titanium (Ti) layer is deposited as an adhesion promoter. Next, 200~nm of Au is deposited on top of the thin Ti layer. The Ti/Au stack is lifted off using NMP, and the wafer is cleaned in solvents and a moderate oxygen plasma in preparation for the next layer.
	
	\subsection{Resonators Etch}
	
	The last step involves patterning and etching the PtSi outlines into LC resonators. It also includes the removal of the sacrificial W layer that was deposited over the PtSi film at the start of the process. To begin, the wafer is once again patterned using the DUV stepper.  The PtSi and W films are etched into resonators with the same etch chemistries that were used in the first layer. The PtSi in this layer is slightly over-etched in order to ensure complete etching in the small inductor gaps (300~nm gaps in the DARKNESS design). After etching, solvents and a moderate oxygen descum are used to remove the remaining photoresist. 
	
\begin{figure*}[ht]
		\centering
		\includegraphics[width=\linewidth]{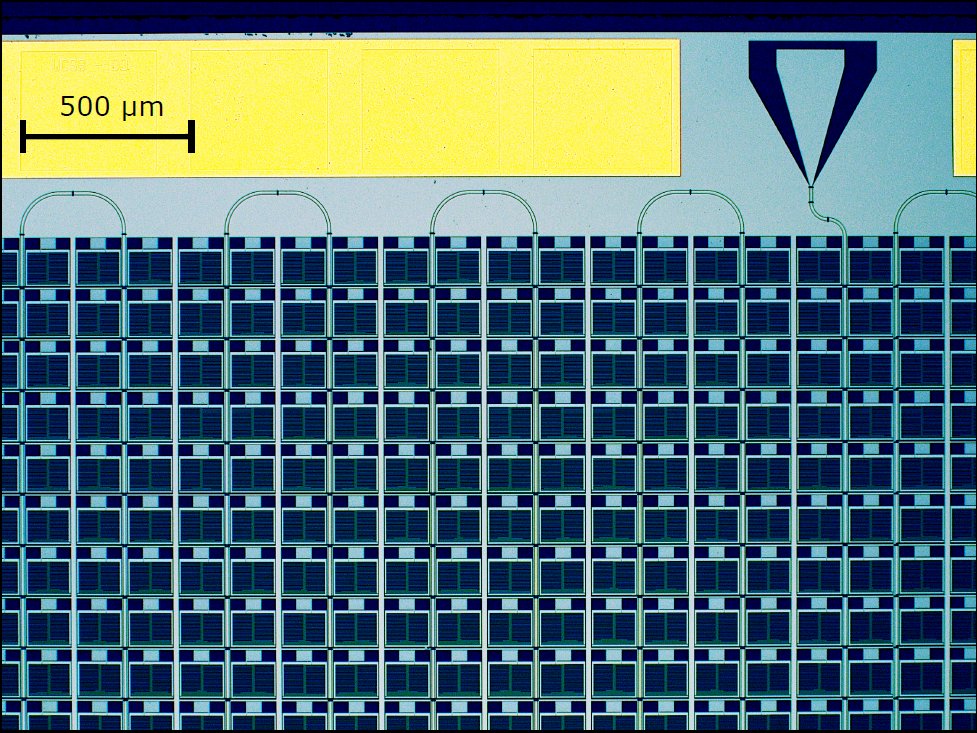}
		\caption{Microscope image of DARKNESS array at the end of fabrication.}
		\label{fig:D3Complete}
	\end{figure*}

	Next, the wafer is prepared for dicing by spinning on a resist layer that protects the structures from impacts and contaminants. The wafer is diced using an ADT 7100 dicing saw.  After dicing, individual devices are solvent cleaned to remove the protective resist. Finally, the W is removed using hydrogen peroxide (H$_2$O$_2$) heated at 50$^{\circ}$C. H$_2$O$_2$ was used because it selectively etches the W and has little to no effect on the other materials used in the process. Fig.~\ref{fig:D3Images}(F) shows the result of this final etch and W removal process.  This completes the fabrication process and at this point the devices can be mounted into boxes and wire bonded. A section of a completed DARKNESS array is shown in Fig.~\ref{fig:D3Complete}.
	
	\section{Characterization}
	
	Upon completion of the fabrication process, a diced array is mounted in a gold-coated copper box using a set of clamps. At the base of this box, dark black Aktar tape is used to prevent reflections of photons that pass through the sensor layer and transparent substrate. Superconducting Al wire bonds are used to connect the microwave transmission lines on the chip to those on the box. Au wire bonds are added to thermally anchor the chip to the box and low temperature device stage. In order to cool the MKID arrays, a Bluefors dilution refrigerator with a base temperature of around 10~mK was used. For the majority of the measurements in this study, the temperature of the device stage was controlled at 100~mK. The results of these measurements are described below.
	
	\subsection{Material Parameters}
	
	The $T_C$ of the PtSi sensor layer was found by performing a DC resistance measurement during a fridge cooldown. Because the etched resonators are too small to wire up for a $T_C$ test, an unused area of PtSi at the side of the wafer is used. The average $T_C$ of the PtSi films was 930~mK, with slight variations from wafer to wafer. Overall, the spatial homogeneity in $T_C$ was very good and similar to that seen in the single layer test devices~\cite{Szypryt2016}.
	
	The sheet inductance, $L_S$, of the PtSi films was determined by measuring the average shift in frequency and backing it out from the simulated frequency and $L_S$ values. This was done using the proportionality relation, $L_{S,m} = L_{S,s} \times f_{0,m} / f_{0,s}$, where the $m$ and $s$ denote the measured and simulated values. Resonant frequencies are found by looking at peaks in the magnitude of the complex transmission with a vector network analyzer (VNA).  The measured values of $L_S$ in more recent wafers were fairly close to the design value of 10.5~pH/$\square$ for 50~nm PtSi films, with little to no average shift in frequency. This is due to the iterative manner in which the wafer $L_S$ is tuned toward the correct value, with small changes in film thickness applied between successive wafers until the inductance is matched.
	
	The W protect layer added an additional variable in tuning the $L_S$. We found that the addition of the protect layer tended to increase the sheet impedance of the final PtSi layer, even after the W was removed with H$_2$O$_2$. This may be attributed to a roughening of the PtSi surface caused by the W removal process. The PtSi film thickness was increased to make up for this effect.
	
	\subsection{Resonator Quality Factors}
	
	The next set of parameters, $Q_i$ and $Q_c$, are calculated by sweeping the frequency space near a resonance and taking in-phase (I) and quadrature (Q) data with a VNA. Each resonator shows up as a loop on the I-Q plane, and this loop can be fit to a resonator model via the procedure given in App.~E of Jiansong Gao's PhD thesis~\cite{Gao2008b} (model given by Eqn.~E.1). The $Q_i$ and $Q_c$ can be extracted from the fitted parameters.
	
\begin{figure*}[ht]
		\centering
		\includegraphics[width=\linewidth]{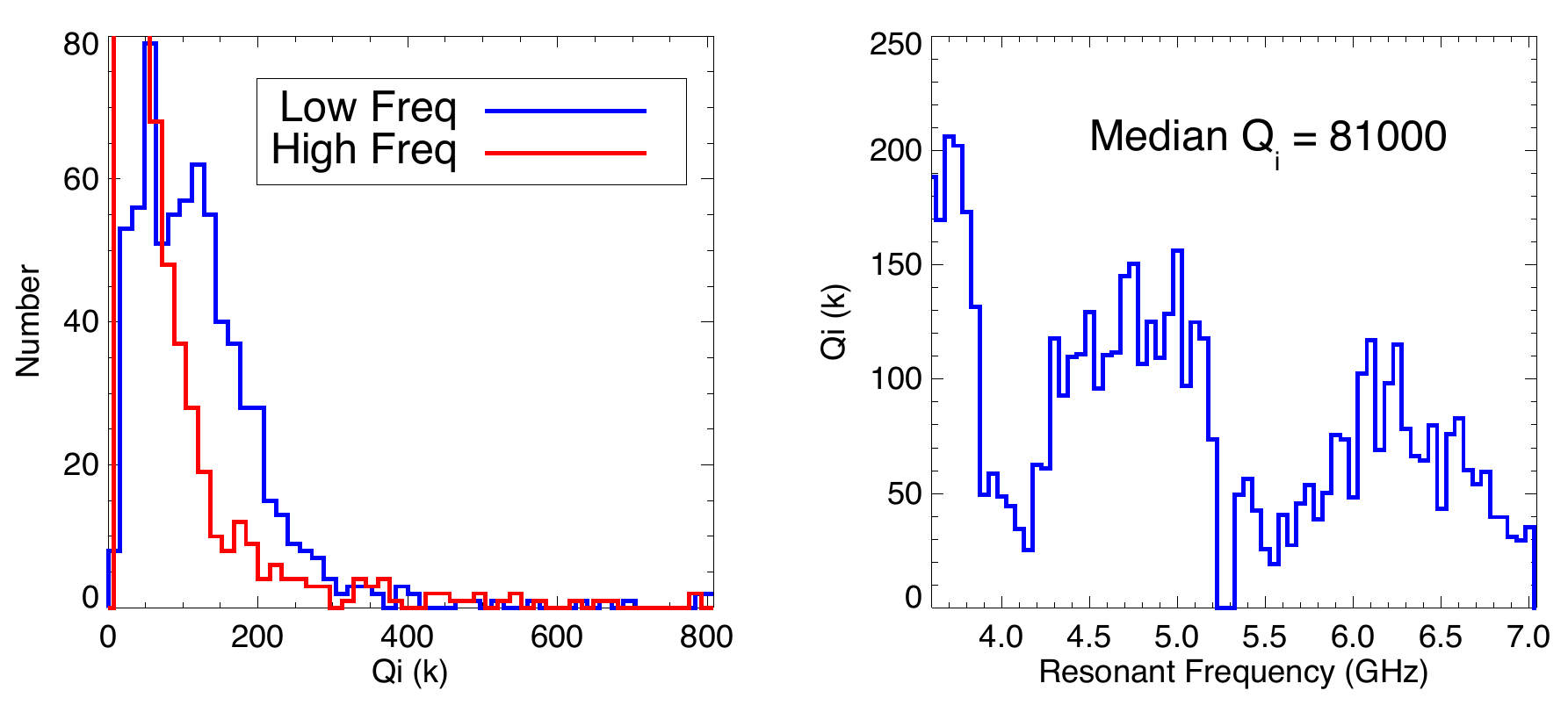}
		\caption{(Left) Histogram of $Q_i$ values for the best performing DARKNESS device, split between the lower and upper halves of the total frequency space. (Right) Plot of median $Q_i$ in a 50 MHz bin versus frequency for the same device. This device has its frequencies shifted away from the designed values due to a PtSi film with higher than expected sheet inductance.}
		\label{fig:DARKNESS_Qi}
	\end{figure*}

\begin{figure*}[ht]
		\centering
		\includegraphics[width=\linewidth]{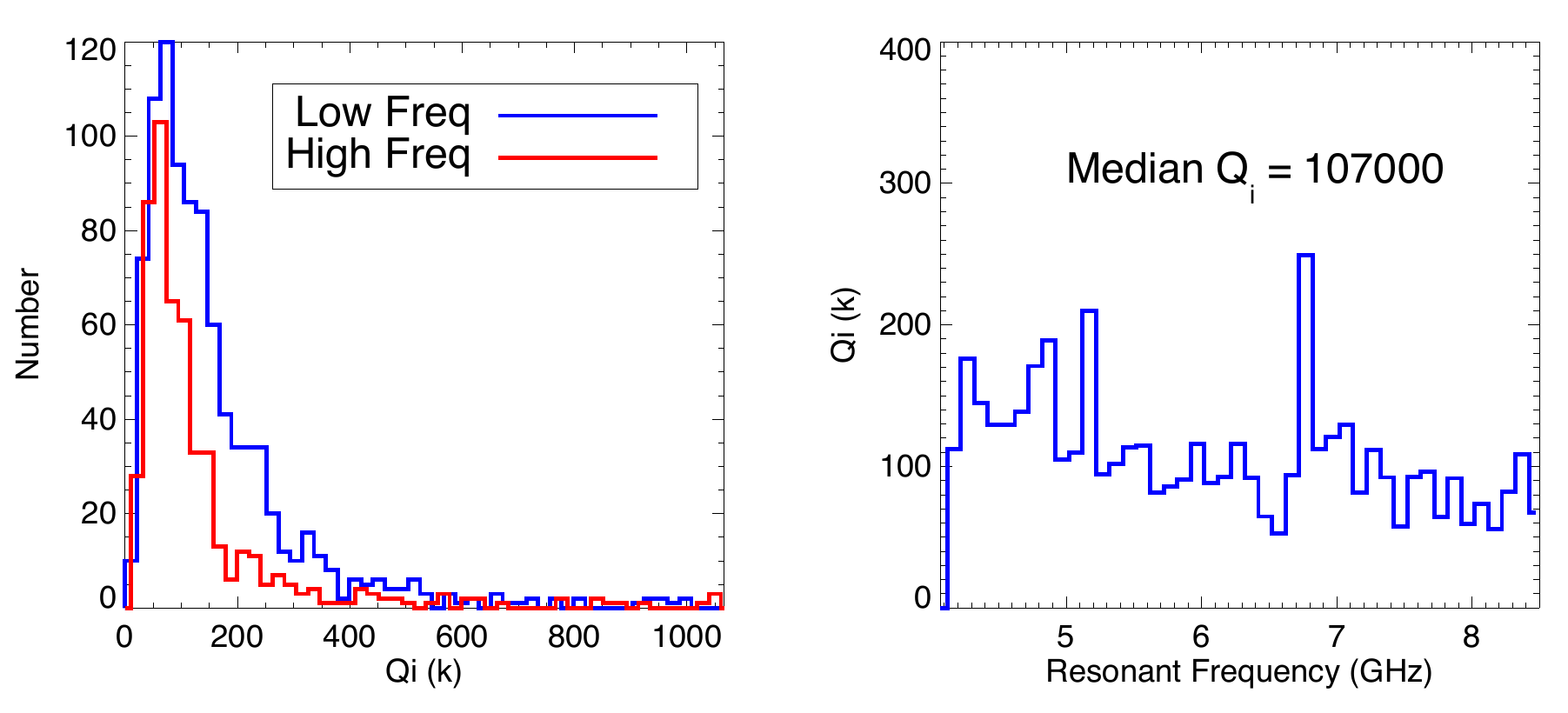}
		\caption{(Left) Histogram of $Q_i$ values for the most recent MEC device, split between the lower and upper halves of the total frequency space. (Right) Plot of median $Q_i$ in a 100 MHz bin versus frequency for the same device. This device had a slight increase in resonant frequencies due to having a sheet inductance smaller than designed.}
		\label{fig:MEC_Qi}
	\end{figure*}

	The median $Q_i$ of the best DARKNESS arrays was about 81,000. Fig.~\ref{fig:DARKNESS_Qi} shows a histogram of $Q_i$ values and a plot of the $Q_i$ values versus frequency for this DARKNESS device. The best MEC arrays had improved $Q_i$, with median values of about 107,000. Fig.~\ref{fig:MEC_Qi} shows an analogous plot for the MEC array.
	
	In terms of $Q_c$, the average value was 28,000 in the above DARKNESS array and 33,000 in the MEC array. These are both in fairly good agreement with the simulated $Q_c$ of 30,000. Small deviations in $Q_c$ such as these are typically not an issue as the readout can vary the amount of power supplied to individual resonators, making detector response more uniform.
	
	\subsection{Spectral Resolution}
	
	The spectral resolution is measured using an analog readout system. Here, the array is illuminated by three sharp laser lines of known wavelength. A histogram of phase shifts is measured, allowing us to correlate phase shifts to energy (or wavelength) across the wavelength band of the instrument.  The histogram is fit to a model of Gaussian peaks, and the widths of the Gaussian peaks are extracted from the model fits. With proper normalization, the width of this Gaussian is proportional to the spread in detected wavelength. The spectral resolution, $R = \lambda / \Delta \lambda$, can then be calculated using this spread.
	
	\begin{figure*}[ht]
		\centering
		\includegraphics[width=0.75\linewidth]{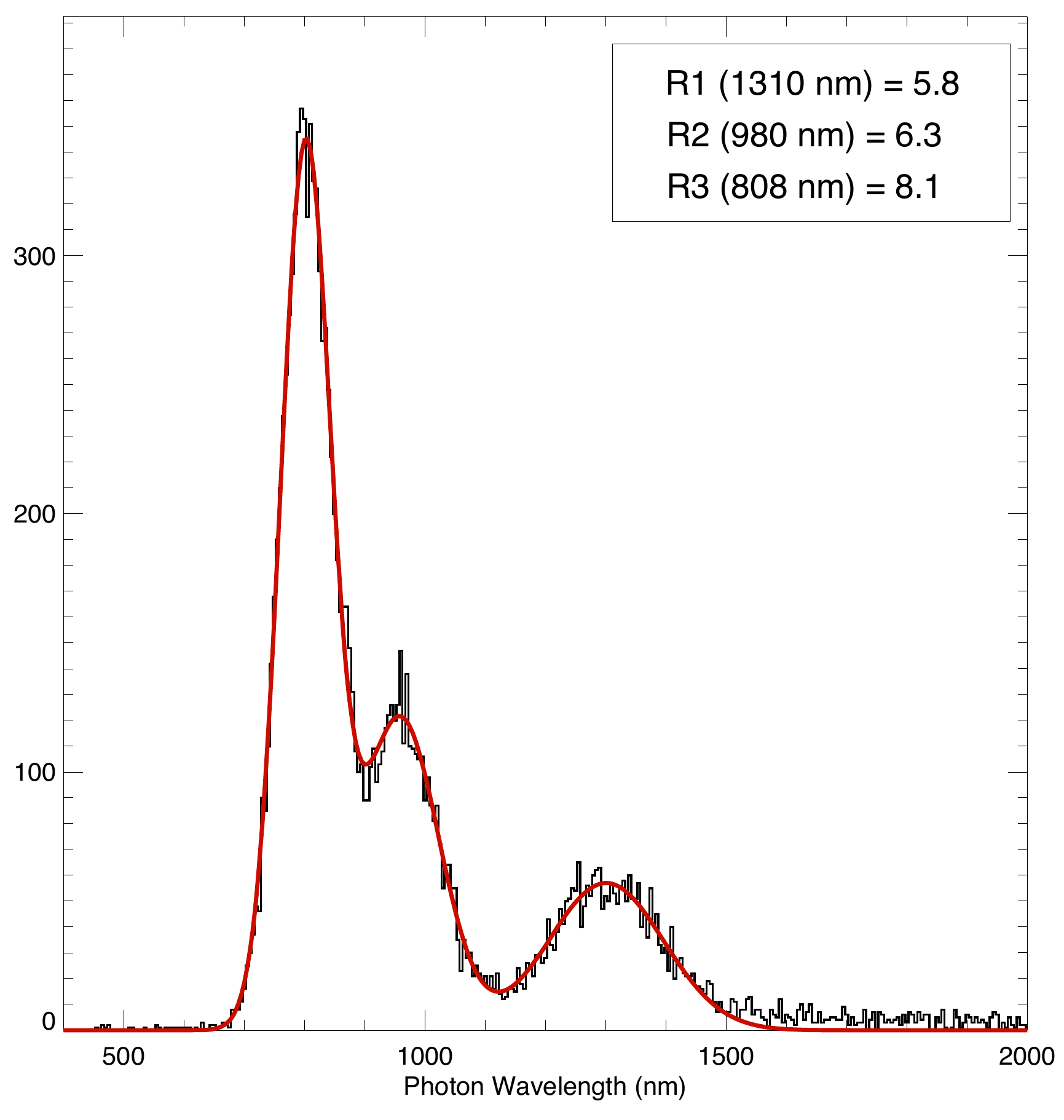}
		\caption{Histogram of photon wavelengths measured using three laser peaks at 1310, 980, and 808~nm. The spectral resolution in the locations of these three peaks was 5.8, 6.3, and 8.1, respectively.}
		\label{fig:energyfit}
	\end{figure*}
	
	DARKNESS and MEC had fairly similar performance in terms of spectral resolution. Both had an spectral resolution of about 8 at 808~nm, with a declining spectral resolution toward higher wavelengths. Fig.~\ref{fig:energyfit} shows a histogram of photon wavelengths (already calibrated from phase shift) for a typical resonator in a large-format PtSi MKID array. The spectral resolution at each laser peak is also included in Fig.~\ref{fig:energyfit}.
	
	\subsection{Yield}
	The yield is measured as the number of unique resonances that can be read out using the room temperature electronics. Resonators are considered overlapping if they are within 500~kHz of another resonator, and these are subtracted from the usable pixel count. The pixel yield is measured on a feedline by feedline basis and is taken as the fraction of good pixels out of the number of total pixels designed on the feedline.
	
	The first DARKNESS arrays had pixel yields of only about 50\%. This was attributed to the rolloff of $Q_i$ at higher frequencies, making those resonators difficult to detect in the readout. This number also reflects pixels that were removed due having low levels of photosensitivity. In the more recent MEC arrays with overall higher values of $Q_i$, the pixel yield is approaching 90\%, after subtracting overlapping resonators. More precise photosensitivity measurements during the MEC commissioning will be required to confirm this pixel yield, and the final pixel count may be reduced.
	
	\section{Discussion}
	
	One of the most striking attributes of these large-format PtSi arrays is the relatively low values of $Q_i$. The best values of $Q_i$ in these arrays were measured in the recent MEC arrays to have values of around $10^5$. In single layer PtSi test resonators, however, the measured $Q_i$ values were on order $10^6$. The difference was attributed to the added complexity of the large-format array process, but it was difficult to pinpoint the fabrication step(s) that was degrading $Q_i$.
	
	One of the first methods used to improve $Q_i$ was to add a layer over the PtSi film to protect it through the subsequent fabrication steps and then remove the protect layer as the very last step. For this method, W was chosen as the protect layer because it could be deposited in-situ after the PtSi, preventing any possible oxidation or contamination of the PtSi surface. The W could also be selectively removed using heated H$_2$O$_2$ with minimal etching of the other materials used in the process. The addition of the W layer raised the $Q_i$ of the resonators by a factor of about two, but this was still too low, causing losses in spectral resolution and pixel yield, especially at high frequencies.
	
	The final resonator etching was also examined as a possible source of $Q_i$ degradation. In the DARKNESS design, the inductor slots were only 300~nm wide, causing them to be etched more slowly than other areas of the resonator. To ensure a complete etching, the total etching time was increased, but this also caused a slight over-etch into the sapphire in the area of the capacitor legs.  The outcome was an increase in $Q_i$ to the levels seen in the best DARKNESS array ($\sim$80,000). To further ensure a complete etching in the MEC design, the inductor slot widths were increased to 500~nm. As was mentioned earlier, the $Q_i$ in the most recent MEC arrays was $\sim$110,000, a considerable improvement from the best DARKNESS arrays. The $Q_i$ is still far lower than the values observed in the single layer test mask and other sources of degradation are being explored.
	
	The performance of the feedline and crossovers was also investigated. As can be seen in Fig.~\ref{fig:DARKNESS_Qi}, there were significant dips in $Q_i$ near 4 and 5.5~GHz. This was attributed to possible transmission line or box standing wave modes not necessarily related to the quality of the resonators. For this reason, high loss microwave absorbers were added to the perimeter of the box to prevent any possible box modes. This did not have any significant effect on the $Q_i$ of the resonators.  In the MEC design, the feedline crossovers connecting the ground plane segments were spaced apart by an arbitrary amount rather than the fixed 150~$\mu$m spacing used in the DARKNESS design. This seemed to significantly reduce standing waves in the transmission line, and the MEC devices did not have these frequency dependent dips in $Q_i$, as can be seen in Fig.~\ref{fig:MEC_Qi}. Different dielectric materials with varying loss tangents underneath the crossovers were also tested. There were no significant variations in the overall transmission between using SiO$_2$ and a-Si, and a-Si was kept as the dielectric material for more recent wafers due to ease of fabrication.
	
	In the most recent devices with average $Q_i > 100,000$, the spectral resolution seems to saturate at R$\sim$8 at 808~nm for a good fraction of the resonators.  At this level, the $Q_i$ does not seem to be the limiting factor for R, indicating the $Q_i$ is getting near the level needed for saturated array performance. Currently, the main factors limiting the spectral resolution are amplifier noise and two-level system noise~\cite{Gao2008a}. These factors require a more significant development effort that is beyond the scope of this work.
	
	\section{Conclusion}
	
	We have fabricated large-format PtSi MKID arrays for the 10,000 pixel DARKNESS instrument and the 20,440 pixel MEC instrument. The resonators on these arrays have measured internal quality factors of up to around $10^5$ and spectral resolutions of 8 at 808~nm. The most recent arrays have potential pixel yields of up to around 90\%. Future work will go toward further increasing the internal quality factors of resonators through a more systematic, layer-by-layer testing procedure. Increases in the internal quality factor will lead to improved pixel yield and slight improvements in spectral resolution. These arrays have already been proven on the sky with the DARKNESS array at Palomar Observatory, and future observations with the improved MEC array at the Subaru Telescope will demonstrate the strengths of optical MKIDs for the direct imaging of exoplanets.
	
	\section*{Funding}
	National Aeronautics and Space Administration (NASA) (NNX16AE98G, NNX13AL70H)
	
	\section*{Acknowledgments}
	This  work  was  supported  by  a  NASA  Space  Technology  Research  Fellowship  (NSTRF). Fabrication of large-format PtSi MKID arrays was done in the UCSB Nanofabrication Facility. Some fabrication was also done at NASA JPL's Microdevices Laboratory (MDL).
	
\end{document}